\documentclass[12pt]{article}
\usepackage[dvips]{graphicx}
\usepackage{amssymb}
\usepackage{amsmath}

\def\beq{\begin{equation}}\def\eeq{\end{equation}}
\def\bea{\begin{eqnarray}}\def\eea{\end{eqnarray}}

\begin{document}

\title{Continuous measurement on a causal set with and without a boundary}
\author{Roman Sverdlov,
\\Department of Mathematics, University of New Mexico} 

\date{June 17, 2020}
\maketitle

\begin{abstract}
The purpose of this paper is two-fold. First, we would like to get rid of common assumption that causal set is bounded and attempt to model its scalar field action under the assumption that it isn't. Secondly, we would like to propose continuous measurement model in this context.   
\end{abstract}

\section{Introduction}

One of the unpleasant features of discrete theories of spacetime is spontaneous violation of principles of relativity. For example, in case of cubic lattice, one can identify preferred spacetime directions with the directions of its edges. For this reason, the majority of theories of quantum gravity use structures other than cubic lattice, where the preferred directions are less obvious. However, they still arise. One of the motivations of causal set theory \cite{Review1, Review2, Review3, Review4, CausalRecent, Surya, Nowhere} is to get rid of these preferred directions. The approach is to replace other discrete structures with simple Poisson process, which respects the principles of relativity. After the Poisson process is performed, one answers yes-or-no question for any given pair of points as to whether or not they are within each other light cone. If it happens that point $x$ is within the past light cone of point $y$ (or, equivalently, point $y$ is within the future light cone of point $x$) one writes $x \prec y$. It has been shown by Hawking \cite{Hawking} and Malament \cite{Malament} that, in a continuum scenario, such partial ordering can recover the metric up to conformal scaling. In light of Poisson process, the conformal scaling can be roughly approximated by the count of points themselves. Thus, there is a rough correspondence between the partial ordering and geometry. The partial ordering is determined in strictly relativistic way, which is the way in which such set respects the principles of relativity. 

This, however, comes with a price. One can check that the Lorentzian $\delta$-neighborhood looks like a vicinity of light cone, and has infinite volume. Consequently, a theory that truly respects principles of relativity will be non-local \cite{CausalNonlocality1, CausalNonlocality2}. In causal set theory this issue is sidestepped by means of an assumption that the entire spacetime has finitely many points which, by default, implies that $\delta$-neighborhood is finite as well. This can either be due to an assertion that the entire causal set is generated by some initial point \cite{Percolation} or by an assertion that it has more general boundary \cite{BoundaryTerms, BoundaryInfluence}. In our opinion, however, both of these assertions compromise the principles of relativity. If we have a boundary, then the shape of the boundary would result in preferred directions. If the boundary consists of a single point, then the preferred direction would be determined by a geodesic connecting the point of our interest to that initial point.  

Within the context of any other theory, this might be non-issue, since the boundary effects are negligible as long as we are sufficiently far away from those boundaries. Within the context of causal set theory, however, this seems less acceptable. After all, one of the goals of causal set theory is to eliminate the microscopic breaking of Lorentz symmetries that are present in other discrete theories. These microscopic violations are negligible too. Therefore, the motivation of causal set theory is not empirical but aesthetic. Empirically, the negligible violations of relativity are inconsequential, but aesthetically they are unpleasant. Thats what we meant by the statement that  the motivation of favoring causal set theory over other discrete theories is aesthetic. This being the case, it seems logically inconsistent to be bothered by microscopic violations of relativity coming from discrete structure, yet not be bothered by macroscopic violations of relativity coming from the shape of the boundaries. Therefore, it would be logical to attempt to remove the boundary within the context of causal set theory -- even if one is willing to have the boundary in other contexts.

 In order to do it, however, we have to find the way of addressing the locality issue. We propose to do that by shifting the focus from the points on the causal set to the edges (where by edge we mean timelike separated pair of points). Suppose we have two timelike edges $(a,b)$ and $(b,c)$. We then put an upper bound on the volume of the region formed by an intersection of future light cone of $a$ and past light cone of $c$. Roughly speaking, this amounts to putting an upper bound on the Lorentzian distance from $a$ to $c$. In light of discreteness, there is statistical lower bound on the distances from $a$ to $b$ and from $b$ to $c$. These lower bounds, together with an upper bound on the distance from $a$ to $c$, implies the upper bound on a hyperbolic angle between $(a,b)$ and $(b,c)$. This, in turn, implies that the number of edges $(a,b)$ that meet this criteria for any given edge $(b,c)$ is finite. 

In terms of quantum field theory, we will alter it in the following way. Instead of attaching fields and Lagrangian densities to the points, we attach them to edges. This would result in Lagrangian densities being local. This, in turn, implies that we can allow causal set to be unbounded while keeping the Lagrangian density finite. However, there is a residual issue. Even if Lagrangian density will be finite and well behaved, the \emph{action} will be infinite since we will be integrating (or, rather, taking discrete sum of) this Lagrangian over unbounded space. A standard approach to this issue is to assume that the field (and, therefore, the Lagrangian density) attenuate at infinity. But then the problem is that the region where the fields are large will be bounded and the shape of its boundary will produce a preferred frame. Once again, this is acceptable in any other context but, in the context of the causal sets, it conflicts with an agenda of trying to preserve relativity at all costs. So we will \emph{refrain} from making the above assumption and look for other ways of addressing the locality issue. 

 We observe that the ultimate prediction of the theory is \emph{not} action but, rather, a set of observables. So our proposed solution is to define a limiting process in such a way that the conditional probability densities of the observables approach finite limit, despite the fact that the action might not do so. However, in order to speak of observables, we need the theory of measurement. We propose to utilize the \emph{continuous measurement model} proposed by Mensky \cite{Mensky1, Mensky2}, Kent \cite{Kent} as well as the author of this paper \cite{Epsilon}. In this paper we will find ways of adapting these models into the causal set framework described above. And we will also explore the new interpretations of quantum measurement that might arise as a result of this. 

\section{Locality and causal sets}  \label{CausalSets} 

\subsection{Review of conventional causal sets} \label{Review} 

Before we proceed, let us review the conventional causal set theory \cite{Review1, Review2, Review3, Review4, CausalRecent}, as done by others. A causal set is a partially ordered set $({\mathcal C}, \prec)$ where the partial ordering $\prec$ is interpreted as lightcone causal relation. That is, $x \prec y$ if and only if $x$ is within past lightcone of $y$ or, equivalently, $y$ is within future light cone of $x$. If \emph{either} $x \prec y$ or $x=y$ we say $x \preceq y$. Thus, for any given $x$, the relation $x \preceq x$ is true, while the relation $x \prec x$ is false. An \emph{interval} or an \emph{Alexandrov set} is defined as 
\begin{equation} I (x,y) = \{z \vert x \preceq z \preceq y \} \end{equation} 
The discretization is postulated through the assertion that $I (x,y)$ is finite for all $x \preceq y$. If $x \prec y$ and $I(x,y)$ happens to be $2$-element set (that is, $I(x,y)= \{x,y \}$) then we write $x \prec^* y$ and say that $x$ is in the \emph{direct past} of $y$ or, equivalently, $y$ is in the \emph{direct future} of $x$. The pair of points $x \prec^* y$ as an \emph{edge}. If either $x \prec^* y$ or $y \prec^* x$ hold, we say that $x$ and $y$ are \emph{direct neighbors}. 

There have been proposals of Delambertians on the causal set \cite{Delambertian1, Delambertian2}. Let us summarize their key aspects. The \emph{n-th layer} of $x$ is defined as 
\begin{equation} L_n (x) = \{y \prec x \vert \sharp I (y,x) = n+1 \} \end{equation}
where $\sharp$ stands for the number of elements. For $n=0$, $L_0 (x)$ is identified with $x$ itself:
\begin{equation} L_0 (x) = \{ x \} \end{equation}
In case of $d$ dimensions, the d'Ambertan is defined as 
\begin{equation} (\Delta \phi)(x) = \frac{1}{l^2} \sum_{k=0}^{n(d)} \bigg(C_{d;k} \sum_{y \in L_k (x)} \phi (y) \bigg) \end{equation}
For $d=2$ and $d=4$ they are given by 
\begin{equation} n(2)= 3 \; , \; C_{2;0}= -2 \; , \; C_{2;1} = 4 \; , \; C_{2;2}= -8 \; , \; C_{2;3}=4 \end{equation}
\begin{equation} n(4) = 4; C_{4;0}= - \frac{4}{\sqrt{6}} \; , \; C_{4;1} = \frac{4}{\sqrt{6}} \; , \; C_{4;2} = - \frac{36}{\sqrt{6}} \; , \; C_{4;3} = \frac{64}{\sqrt{6}} \; , \; C_{4;4} =- \frac{32}{\sqrt{6}} \end{equation} 
In light of the non-locality of Lorentzian neighborhoods, the sets $L_k (x)$, $k \geq 1$, have infinitely many elements -- and most of these elements are arbitrarily far away coordinate-wise. They are on the ``tails" of Lorentzian neighborhood that fills the vicinity of lightcone. It has been speculated, and confirmed numerically, that the contributions from the ``tails" cancel each other out, leaving the result that roughly approximates the well known Lagrangian. However, this statement is not well defined \emph{unless} causal set is finite. After all, Lorentz invariance tells us that there is no reason to expect the contributions from the ``tails" to get smaller which, in turn, implies that the infinite series would diverge. Therefore, given that our agenda is to allow causal set to be infinite, we have to modify this somehow. 

\subsection{Replacement of points with edges} \label{Edges}

In order to remove the above infinity, we observe that, while the set of points is non-local, the set of \emph{edges} is local. Therefore, we propose to restore locality by replacing points with edges in the above theory. In order to make this statement precise, let us define a partial ordering on the set of \emph{edges}. For any given $\Lambda \in \mathbb{N}$ we will define a relation between the edge $x \prec^* y$ and an edge $y \prec^* z$, which we denote by $(x \prec^* y) \rightarrow^*_{\Lambda} (y \prec^* z)$, by a statement that the number of elements of $I(x,z)$ is less than or equal to $\Lambda$. We then transitively extend it to define a relation $(x \prec^* y) \rightarrow_{\Lambda} (z \prec^* w)$ by the statement that one can find the sequence $(x \prec^* y) \rightarrow^*_{\Lambda} (y \prec^* \xi_1) \rightarrow^*_{\Lambda} \cdots \rightarrow^*_{\Lambda} (\xi_{n-1} \prec^* z) \rightarrow^*_{\Lambda} (z \prec^* w)$. One can show that, for any given edge $y \prec^* x$, the number of edges $z \prec^* y$ satisfying the relation $(z \prec^* y) \rightarrow^*_{\Lambda} (y \prec^* x)$ is finite (See Section \ref{ExpectationValueEdges}), although it approaches infinity in the limit of $\Lambda \rightarrow \infty$. Suppose we assume that 
\begin{equation} l \Lambda = l_0 \label{ConstraintEdges} \end{equation}
where $l$ is the average distance between direct neighbors and $l_0$ is agreed-upon constant, such as Planck lenth. In this case, in the limit $\Lambda \rightarrow \infty$, the edge becomes more and more similar to the point and, at the same time, the number of its edge-neighbors gets larger and larger. Therefore, we will assume that $\Lambda$ is very large but finite. The fact that it is large will enable us to approximate the results we discussed in Section \ref{Review}; the fact that it is still finite will enable us to say that the sum is mathematically well defined. We will think of $\Lambda$ as integer-valued physical constant, that simply \emph{happened} to be very large. 

 We then define the intervals on the set of edges as
\begin{equation} I_{\Lambda} \Big((x \prec^*y), (z \prec^* w)\Big) = \nonumber \end{equation}
\begin{equation} = \Big\{ (x \prec^* y), (z \prec^* w) \Big\} \cup \Big\{ (\xi \prec^* \eta) \Big\vert (x \prec^* y) \rightarrow_{\Lambda} (\xi \prec^* \eta) \rightarrow_{\Lambda} (z \prec^* w) \Big\} \end{equation} 
We then define layers as follows:
\begin{equation} L_{\Lambda, n} (x \prec^* y) = \bigg\{ (z \prec^* w) \bigg\vert \sharp I \Big((z \prec^* w), (x \prec^* y) \Big) = n+1 \bigg\} \end{equation}
We then attach field to edges instead of attaching it to points. That is, we replace $\phi (x)$ with $\phi (x \prec^* y)$. Likewise, we attach d'Ambertan to edges as well, and define it as 
\begin{equation} (\Delta \phi)_{\Lambda} (x \prec^* y) = \frac{1}{l^2} \sum_{k=0}^{n(d)} \bigg(C_{d;k} \sum_{z \prec^*w \in L_{\Lambda,k} (x \prec^* y)} \phi (z \prec^* w) \bigg) \label{EdgeDambertan} \end{equation}
Our goal is for the above result to approximate the result of conventional causal set theory since we already know from the numerical studies that in the conventional causal set theory we obtain d'Ambertan. Now, in the conventional situation, the number of terms is still finite, due to the boundary, yet it is very large. Similarly, the assertion that $\Lambda$ is very large coupled with Eq \ref{ConstraintEdges} accomplishes the same thing -- without having to assume the existence of the boundaries. As stated earlier, we will treat $\Lambda$ as a physical constant; we simply don't know its value, other than the fact that its value is very large. 

\subsection{Expectation value of the number of edge neighbors} \label{ExpectationValueEdges}

In the previous section, we made an assertion that the number of edge-neighbors is finite. Let us now prove this assertion by explicitly calculating the number of edge-neighbors of a given edge. 

In our calculation will be repeatedly using the expression for the area of the $d-2$-dimensional sphere in $d-1$ dimensional Eucledian space, the volume of $d-1$ dimensional ball, again in $d-1$ dimensional space, and the volume of Alexandrov set in $d$-dimensional space. As far as the sphere and the ball, these formulas are well known: 
\begin{equation} a_{d-2} = 2 \frac{\pi^{(d-1)/2}}{\Gamma (\frac{d-1}{2})} \end{equation}
\begin{equation} v_{d-1} = \frac{2}{d-1} \frac{\pi^{(d-1)/2}}{\Gamma(\frac{d-1}{2})} \end{equation}
Let us now compute the formula for an Alexandrov set. Suppose we have an Alexandrov set with distance between the poles being equal to $\tau$. Then the volume of the Alexandrov set will be equal to twice the volume of half of Alexandrov set. And half of Alexandrov set will be composed of spheres whose ``height" is $t \in (0, \tau/2)$ and whose radius is $t$. Therefore, 
\begin{equation} V = 2 \int_0^{\tau/2} \frac{2}{d-1} \frac{\pi^{(d-1)/2}}{\Gamma(\frac{d-1}{2})} t^{d-1} dt = 2  \frac{2}{d-1} \frac{\pi^{(d-1)/2}}{\Gamma(\frac{d-1}{2})} \frac{1}{d} \bigg(\frac{\tau}{2} \bigg)^d = \nonumber \end{equation}
\begin{equation} =  \frac{\pi^{(d-1)/2} \tau^d}{d(d-1)2^{d-2} \Gamma(\frac{d-1}{2})} = k_d \tau^d \end{equation} 
where 
\begin{equation} k_d =\frac{\pi^{(d-1)/2}}{d(d-1)2^{d-2} \Gamma(\frac{d-1}{2})}  \end{equation}
We are now ready to return to the calculation of the main result. Suppose we already know that there are points $p$ and $q$ in the scatter located at 
\begin{equation} p = (- \tau, 0, \cdots, 0) = (- \tau, \vec{0}) \end{equation}
\begin{equation} q = (0, 0, \cdots, 0) = (0, \vec{0}) \end{equation}
Let $\Delta$ be some region of very small volume $\delta v$ around point $x$. We would like to find the expectation value of the number of points $r \in \Delta$ that would obey $(p,q) \prec_{\Lambda} (q,r)$. The expectation number of the total number of points that fall into that region is $\rho \delta V$, and the probability of each of these points obeying the above relation is approximately $\frac{e^{- \rho V(p,x)} (\rho V(p,x))^{\Lambda}}{\Lambda!}$ where $V(p,x)$ is the volume of $I (p,x)$. Therefore, the expectation value of the number of points $r \in \Delta$ obeying $(p,q) \prec_{\Lambda} (q,r)$ is approximately $\frac{e^{- \rho V(p,x)} (\rho V(p,x))^{\Lambda}}{\Lambda!} \rho \delta v$. One can show that the expectation value is additive. Therefore, if we relax the condition regarding the location of $r$ and only retain the condition that $(p,q) \prec_{\Lambda} (q,r)$, then we can partition the future part of lightcone of $q$ into such small regions and add these expectation values. This will result in Riemann sum. Finally, to get an exact answer we take the limit of their volumes approaching $0$ and this will result in the integral. Thus, 
\begin{equation} \mathbb{E} (\sharp \{r \vert (p, q) \prec_{\Lambda} (q,r) \}) = \int \frac{e^{- \rho V(p,x)} (\rho V(p,x))^{\Lambda}}{\Lambda!} \rho dv  \label{Expect} \end{equation} 
Now consider the hypersurface $\vert x -p \vert = \xi$.  Let $\theta (x)$ be the hyperbolic angle between the segment $px$ and segment $pq$ at which this hypersurface will intersect the light cone of $q$. Then we have
\begin{equation} 0 = (- \tau + \xi \cosh \theta (\xi))^2 - (\xi \sinh \theta (\xi))^2 = \nonumber \end{equation}
\begin{equation} = \tau^2 - 2 \tau \xi \cosh \theta (\xi) + \xi^2 \cosh^2 \theta(\xi) - \xi^2 \sinh^2 \theta(\xi) = \tau^2 - 2 \tau \xi \cosh \theta (\xi) + \xi^2 \end{equation} 
from which we obtain 
\begin{equation} \theta (\xi) = \cosh^{-1} \frac{\tau^2 + \xi^2}{2 \tau \xi} \end{equation}
One can show that the expression for $\cosh^{-1}$ is 
\begin{equation} \cosh^{-1} \lambda = \ln (\lambda + \sqrt{\lambda^2-1}) \end{equation}
and, therefore, 
\begin{equation} \theta (\xi) = \ln \bigg(\frac{\tau^2 + \xi^2}{2 \tau \xi} + \sqrt{\bigg(\frac{\tau^2 + \xi^2}{2 \tau \xi} \bigg)^2-1} \bigg) = \nonumber \end{equation}
\begin{equation} = \ln \bigg(\frac{\tau^2 + \xi^2}{2 \tau \xi} + \frac{\sqrt{(\tau^2 + \xi^2)^2 - 4 \tau^2 \xi^2}}{2 \tau \xi} \bigg) = \ln \bigg(\frac{\tau^2 + \xi^2}{2 \tau \xi} + \frac{\sqrt{\tau^4 + 2 \tau^2 \xi^2 + \xi^4 - 4 \tau^2 \xi^2}}{2 \tau \xi} \bigg) = \nonumber \end{equation}
\begin{equation} = \ln \bigg(\frac{\tau^2 + \xi^2}{2 \tau \xi}  + \frac{\sqrt{\tau^4 -2 \tau^2 \xi^2 + \xi^4}}{2 \tau \xi} \bigg) = \ln \bigg(\frac{\tau^2 + \xi^2}{2 \tau \xi}  +  \frac{\sqrt{(\xi^2- \tau^2)^2}}{2 \tau \xi} \bigg) = \nonumber \end{equation}
\begin{equation} =  \ln \bigg(\frac{\tau^2 + \xi^2}{2 \tau \xi}  +  \frac{\xi^2- \tau^2}{2 \tau \xi} \bigg) = \ln \frac{\tau^2 + \xi^2 + \xi^2- \tau^2}{2 \tau \xi} = \ln \frac{2 \xi^2}{2 \tau \xi} = \ln \frac{\xi}{\tau} \end{equation} 
where in the first equal sign on the last line we have used the fact that $\xi \geq \tau$ and, therefore, $\sqrt{(\xi^2- \tau^2)^2} = \xi^2- \tau^2$ as opposed to $\tau^2 - \xi^2$. Let $A (\xi)$ be the area of the surface bounded by $\theta (\xi)$. It is given by
\begin{equation} A(\xi) = \int_0^{\theta (\xi)} a_{d-2} (\xi \sinh \theta)^{d-2} \xi d \theta = a_{d-2} \xi^{d-1} \int_0^{\theta (\xi)} \sinh^{d-2} \theta d \theta = \nonumber \end{equation}
\begin{equation} = a_{d-2} \xi^{d-1} \int_0^{\theta (\xi)} \bigg(\frac{e^{\theta} + e^{- \theta}}{2} \bigg)^{d-2} d \theta = \frac{a_{d-2} \xi^{d-1}}{2^{d-2}} \int_0^{\theta (\xi)} \sum_{k=0}^{d-2} {d-2 \choose k} (e^{\theta})^{d-2-k} (e^{- \theta})^k d \theta = \nonumber \end{equation}
\begin{equation} = \frac{a_{d-2} \xi^{d-1}}{2^{d-2}} \int_0^{\theta (\xi)} \sum_{k=0}^{d-2} {d-2 \choose k} e^{(d-2-2k) \theta} d \theta = \frac{a_{d-2} \xi^{d-1}}{2^{d-2}}  \sum_{k=0}^{d-2} {d-2 \choose k} \frac{e^{(d-2-2k) \theta (\xi)} -1}{d-2-2k}  \end{equation} 
Now if we substitute $\theta (\xi)= \ln (\xi/ \tau)$ we obtain 
\begin{equation} e^{(d-2-2k) \theta (\xi)} = (e^{\theta (\xi)})^{d-2-2k} = (e^{\ln (\xi/ \tau)})^{d-2-2k} = \bigg(\frac{\xi}{\tau} \bigg)^{d-2-2k} \end{equation} 
and, therefore
\begin{equation} A (\xi) = \frac{a_{d-2} \xi^{d-1}}{2^{d-2}}  \sum_{k=0}^{d-2} {d-2 \choose k} \frac{e^{(d-2-2k) \theta (\xi)} -1}{d-2-2k} =  \frac{a_{d-2} \xi^{d-1}}{2^{d-2}}  \sum_{k=0}^{d-2} {d-2 \choose k} \frac{(\frac{\xi}{\tau})^{d-2-2k} -1}{d-2-2k} = \nonumber \end{equation} 
\begin{equation} =  \frac{a_{d-2}}{2^{d-2}}  \sum_{k=0}^{d-2} {d-2 \choose k} \frac{\frac{ \xi^{d-1} \xi^{d-2-2k}}{\tau^{d-2-2k}} -\xi^{d-1}}{d-2-2k} =  \frac{a_{d-2}}{2^{d-2}}  \sum_{k=0}^{d-2} {d-2 \choose k} \frac{\frac{ \xi^{2d-3-2k}}{\tau^{d-2-2k}} -\xi^{d-1}}{d-2-2k}  \label{AofXi} \end{equation} 
Now, Eq \ref{Expect} tells us 
\begin{equation} \mathbb{E} (\sharp \{r \vert (p, q) \prec_{\Lambda} (q,r) \}) = \int \frac{e^{- \rho V(p,x)} (\rho V(p,x))^{\Lambda}}{\Lambda!} \rho dv  \end{equation}  
On a hypersurface $\tau (p,x) = \xi$ we have 
\begin{equation} V(p,x) = k_d \tau^d (p,x) = k_d \xi^d \end{equation}
and, therefore, the expression under the integral is a function of $\xi$ alone: 
\begin{equation} \frac{e^{- \rho V(p,x)} (\rho V(p,x))^{\Lambda}}{\Lambda!}  =  \frac{e^{- \rho k_d \xi^d} (\rho k_d \xi^d)^{\Lambda}}{\Lambda!}  \end{equation}
This being the case, we can replace $dv$ with the volume of the slice produced by the above hypersurface with thickness $d \xi$:
\begin{equation} \rho dv = \rho A (\xi) d \xi \end{equation}
and therefore 
\begin{equation} \mathbb{E} (\sharp \{r \vert (p, q) \prec_{\Lambda} (q,r) \}) = \int \frac{e^{- \rho k_d \xi^d} (\rho k_d \xi^d)^{\Lambda}}{\Lambda!}  \rho A (\xi) d \xi \end{equation}
By substituting the expression for $A (\xi)$ given in Eq \ref{AofXi} we obtain 
\begin{equation} \mathbb{E} (\sharp \{r \vert (p, q) \prec_{\Lambda} (q,r) \}) = \int \frac{e^{- \rho k_d \xi^d} (\rho k_d \xi^d)^{\Lambda}}{\Lambda!}  \rho \frac{a_{d-2}}{2^{d-2}}  \sum_{k=0}^{d-2} {d-2 \choose k} \frac{\frac{ \xi^{2d-3-2k}}{\tau^{d-2-2k}} -\xi^{d-1}}{d-2-2k}  d \xi = \nonumber \end{equation}
\begin{equation} = \frac{a_{d-2}}{\Lambda!2^{d-2}} \sum_{k=0}^{d-2} \frac{{d-2 \choose k}}{d-2-2k}  \bigg(\frac{1}{\tau^{d-2-2k}} \int e^{- \rho k_d \xi^d} (\rho k_d \xi^d)^{\Lambda} \xi^{2d-3-2k} \rho d \xi - \int e^{- \rho k_d \xi^d} (\rho k_d \xi^d)^{\Lambda} \xi^{d-1}  \rho d \xi \bigg) \end{equation} 
Lets change the variables to 
\begin{equation} \eta = \rho k_d \xi^d \end{equation} 
Then we have 
\begin{equation} \xi^{d-1}\rho d \xi = \frac{\rho}{d} d \xi^d = \frac{\rho}{d} d \frac{\eta}{\rho k_d} = \frac{d \eta}{k_d d} \end{equation} 
and, therefore, 
\begin{equation} \int e^{- \rho k_d \xi^d} (\rho k_d \xi^d)^{\Lambda} \xi^{d-1}  \rho d \xi = \int e^{- \eta} \eta^{\Lambda} \frac{d \eta}{k_dd} = \frac{\Gamma(n+1)}{k_dd} = \frac{\Lambda!}{k_dd}  \end{equation} 
On the other hand, 
\begin{equation} \xi^{2d-3-2k} \rho d \xi = \xi^{d-2-2k} \xi^{d-1} \rho d \xi = \bigg(\frac{\eta}{\rho k_d} \bigg)^{(d-2-2k)/d}  \frac{d \eta}{k_d d} \end{equation} 
and therefore 
\begin{equation} \int e^{- \rho k_d \xi^d} (\rho k_d \xi^d)^{\Lambda} \xi^{2d-3-2k} \rho d \xi = \int e^{- \eta} \eta^{\Lambda} \bigg(\frac{\eta}{\rho k_d} \bigg)^{(d-2-2k)/d}  \frac{d \eta}{k_d d} = \nonumber \end{equation} 
\begin{equation} = \frac{1}{\rho^{1- \frac{2+2k}{d}} k_d^{2- \frac{2+2k}{d}} d} \int e^{- \eta} \eta^{\Lambda+1- \frac{2+2k}{d}} d \eta =  \frac{\Gamma (n+2- \frac{2+2k}{d})}{\rho^{1- \frac{2+2k}{d}} k_d^{2- \frac{2+2k}{d}} d}  \end{equation} 
We thus obtain
\begin{equation} \mathbb{E} (\sharp \{r \vert (p,q) \prec_{\Lambda} (q,r) \}) = \frac{a_{d-2}}{\Lambda!2^{d-2}} \sum_{k=0}^{d-2} \frac{{d-2 \choose k}}{d-2-2k}  \bigg(\frac{1}{\tau^{d-2-2k}}   \frac{\Gamma (n+2- \frac{2+2k}{d})}{\rho^{1- \frac{2+2k}{d}} k_d^{2- \frac{2+2k}{d}} d} -  \frac{\Lambda!}{k_dd} \bigg) = \nonumber \end{equation} 
\begin{equation} = \frac{a_{d-2}}{\Lambda!2^{d-2}} \sum_{k=0}^{d-2} \frac{{d-2 \choose k}}{d-2-2k}  \bigg(\frac{\Gamma (n+2- \frac{2+2k}{d})}{d(\rho^{1/d}\tau)^{d-2-2k}}   \frac{1}{ k_d^{2- \frac{2+2k}{d}} } -  \frac{\Lambda!}{k_dd} \bigg)  \end{equation} 
By substituting
\begin{equation} a_{d-2} = 2 \frac{\pi^{(d-1)/2}}{\Gamma(\frac{d-1}{2})} \end{equation} 
\begin{equation} k_d =\frac{\pi^{(d-1)/2}}{d(d-1)2^{d-2} \Gamma(\frac{d-1}{2})}  \end{equation}
this becomes 
\begin{equation} \mathbb{E} (\sharp \{r \vert (p,q) \prec_{\Lambda} (q,r) \}) =   \end{equation}
\begin{equation} =  \frac{\pi^{(d-1)/2}}{\Lambda!2^{d-3} \Gamma(\frac{d-1}{2})}  \sum_{k=0}^{d-2} \frac{{d-2 \choose k}}{d-2-2k}  \bigg(\frac{\Gamma (n+2- \frac{2+2k}{d})}{d(\rho^{1/d}\tau)^{d-2-2k}}   \bigg(\frac{d(d-1)2^{d-2} \Gamma(\frac{d-1}{2})}{\pi^{(d-1)/2}}\bigg)^{2- \frac{2+2k}{d}} -  \frac{\Lambda!}{k_dd} \bigg) \nonumber \end{equation}
At this point one would have to compute the expectation value of $\frac{1}{\tau^{d-2-2k}}$. This would in turn require a mathematically rigorous definition of expectation value over infinite set, that would end up being surprisingly difficult. But, for the purposes of this paper, the only thing we would like to know is that the expectation value of the number of edge-neighbors is \emph{finite} (and it doesn't really matter what that finite number happens to be). The above formula shows that it is, in fact, finite. 

\section{Continuous measurement model}

\subsection{Continuum version} \label{EpsilonContinuum}

In the present sub-section we will digress and summarize the continuum measurement model proposed in \cite{Epsilon} in the continuum case. Therefore, for the purposes of this sub-section, the scalar field is, once again, a function of points as opposed to pairs of points. Then, in the next sub-section, we will modify our results to accommodate the pairs of points.

Consider, for simplicity, a quantum field theory with a single scalar field, $\phi$, that has Lagrangian 
\begin{equation} {\cal L} (\phi;x) = \frac{1}{2} \partial^{\mu} \phi \partial_{\mu} \phi - \frac{m^2}{2} \phi^2 - \frac{\lambda}{4!} \phi^4 \label{Phi4} \end{equation}
that corresponds to an action 
\begin{equation} S (\phi) = \int d^4 x {\cal L} (\phi; x) \end{equation}
According to weighted path integral model of quantum measurement, there is an observable\footnote{While in current paper we use the term observabe and letter $\phi_{ob}$, in \cite{Epsilon} we used the word ``classical" and letter $\phi_{cl}$ to mean the same thing. What we mean by ``classical" is strictly an ontology. Quantum mechanical entities take multiple trajectories at the same time while classical entities take one single trajectory. We did \emph{not} mean to say that any of the laws of classical physics hold (they do not!) However, upon subsequent conversations with other scientists, we noticed that when people hear the word ``classical" they often understand it to mean as if we claim the laws of classical physics hold, which is not what we are trying to say. That is why in this paper we decided to replace ``classical" with ``observable". Accordingly, we replace $\phi_{cl}$ with $\phi_{ob}$} field $\phi_{ob}$ that co-exists with the field that has quantum mechanical ontology, $\phi$. On the one hand, we take path integral over different trajectories of $\phi$, which results in interference effects. On the other hand, as far as $\phi_{ob}$ is concerned, it will take one single trajectory, without any interference. But that trajectory is not determined until we make a ``measurement". The ``measurement" takes a form of a single snapshot of the entire spacetime history. Thus, we do not have more than one measurement which renders the causality paradoxes coming from two or more measurements irrelevant. The probability density that the result of the measurement of $\phi_{ob}$ results in a specific trajectory is given by
\begin{equation} \rho (\phi_{ob}) = \frac{\vert Z (\phi_{ob}) \vert^2}{\int[{\cal D} \phi_{ob}] \vert Z (\phi_{ob}) \vert^2} \label{BornBounded}\end{equation}
where $[{\cal D} \phi_{ob}]$ is the measure on the space of trajectories that is heuristically given by 
\begin{equation} [{\cal D} \phi_{ob}] = \prod_x d\phi (x) \label{FeynmannMeasure} \end{equation} 
and $Z(\phi_{ob})$ is given by \emph{weighted path integral}
\begin{equation} Z(\phi_{ob}) = \int [{\cal D} \phi] w(\phi, \phi_{ob}) e^{iS(\phi)} \end{equation}
and $w (\phi, \phi_{ob})$ is a \emph{weight function} given by 
\begin{equation} w (\phi, \phi_{ob}) = \exp \bigg(- \frac{\alpha}{2} \int d^4 x (\phi(x) - \phi_{ob} (x))^2 \bigg) \label{WeightFunction} \end{equation} 
This can be restated as 
\begin{equation} Z(\phi_{ob}) = \int [{\cal D} \phi] e^{iS_{\alpha} (\phi)} \end{equation}
where
\begin{equation} S_{\alpha} (\phi) = S(\phi) + \frac{i \alpha}{2} \int d^4 x (\phi(x) - \phi_{ob} (x))^2 \end{equation}
This corresponds to the Lagrangian 
\begin{equation} {\mathcal L}_{\alpha} (\phi) = {\mathcal L} (\phi) + \frac{i \alpha}{2} (\phi(x) - \phi_{ob} (x))^2 \end{equation}
By substituting Eq \ref{Phi4} we obtain 
\begin{equation}  {\mathcal L}_{\alpha} (\phi) = \frac{1}{2} \partial^{\mu} \phi \partial_{\mu} \phi - \frac{m^2}{2} \phi^2 - \frac{\lambda}{4!} \phi^4 + \frac{i \alpha}{2} (\phi(x) - \phi_{ob} (x))^2 \end{equation}
which can be re-expressed as 
\begin{equation}   {\mathcal L}_{\alpha} (\phi) = \frac{1}{2} \partial^{\mu} \phi \partial_{\mu} \phi - \frac{m^2- i \alpha}{2} \phi^2 - \frac{\lambda}{4!} \phi^4  - i \alpha \phi_{ob}  \phi + \frac{i \alpha}{2} \phi_{ob}^2 \end{equation}
This can be further rewritten as
\begin{equation}   {\mathcal L}_{\alpha} (\phi) = \frac{1}{2} \partial^{\mu} \phi \partial_{\mu} \phi - \frac{m_{\alpha}^2}{2} \phi^2 - \frac{\lambda}{4!} \phi^4 +J_{\alpha} \phi + \frac{i \alpha}{2} \phi_{ob}^2 \end{equation} 
where
\begin{equation} m_{\alpha}^2 = m^2 - i \alpha \end{equation}
\begin{equation} J_{\alpha} (x) = - i \alpha \phi_{ob} (x) \end{equation}
We claim that $J_{\alpha}$ corresponds to the sources and sinks in the conventional quantum field theory (which is why we chose to use the letter $J$). And, indeed, the conventional quantum field theory calculations predict locations of sources and sinks -- which means that they predict the trajectory of $J$, except that they make an extra assumption that $J$ is a sum of $\delta$-functions -- which they don't have to make. In our case we are predicting the trajectory of $\phi_{ob}$. So it makes sense that $\phi_{ob}$ and $J$ are related. The ``new" mass $m_{\alpha}$ leads to the propagator
\begin{equation} \frac{1}{p^2 -m_{\alpha}^2} = \frac{1}{p^2-m^2+i \alpha} \end{equation}
If we take 
\begin{equation} \alpha = \epsilon \end{equation}
our propagator will become 
\begin{equation} \frac{1}{p^2-m^2+i \epsilon} \end{equation}
that we are familiar with, except that $i \epsilon$ is no longer just a trick to avoid the poles but, instead, it is a real physical parameter that comes from weighted path integral. Roughly speaking, the field variations are classically observable if they are much greater than some power of the inverse of $\epsilon$. The assumption that $\epsilon$ is infinitesimal is equivalent to the assumption that none of the field strengths are classically observable. And, indeed, both assumptions are made in the conventional quantum field theory. If we stick with the assumption that $\epsilon$ is finite, then $e^{imt}$ becomes
\begin{equation} e^{-im_{\alpha}t} = \exp \Big(-i \sqrt{m^2 -i \epsilon} t \Big) = \exp \bigg(-it \bigg(m - \frac{i \epsilon}{2m} \bigg) \bigg) + 0 (\epsilon^2) = \nonumber \end{equation}
\begin{equation} = \exp \bigg(-imt  - \frac{\epsilon t}{2m} \bigg) + 0 (\epsilon^2) = e^{-imt} e^{\epsilon t/2m} + 0 (\epsilon^2) \label{ContinuumAttenuation} \end{equation}
The extra factor of $e^{- \epsilon t/ 2m}$ implies that the signals attenuate and their influence is negligible when $t \gg \epsilon^{-1}$. We will use this fact in order to utilize locality for our purposes.

\subsection{Causal set version with the boundary}

Let us now discretized what we did in Section \ref{EpsilonContinuum}. In this section, we will do the discretization under the assumption that the boundary is present. Then, in the next section, we will explore how to generalize it to the cases when the boundary is absent. We will rewrite Eq \ref{Phi4} as 
\begin{equation} {\mathcal L}_{\Lambda} (\phi; x) = - \frac{1}{2} \phi \Delta_{\Lambda} \phi - \frac{m^2}{2} \phi^2 - \frac{\lambda}{4!} \phi^4 \end{equation}
As explained in Sec \ref{Edges}, we replace $\phi (x)$ with $\phi (x \prec^* y)$. Similarly, we will replace ${\cal L} (\phi; x)$ with ${\cal L} (\phi; x \prec^* y)$. By substituting Eq \ref{EdgeDambertan} we obtain 
\begin{equation} {\cal L}_{\Lambda} (\phi; x \prec^* y) = - \frac{m^2}{2} \phi^2 (x \prec^* y) - \frac{\lambda}{4!} \phi^4 (x \prec^* y) - \nonumber \end{equation}
\begin{equation} -  \frac{1}{2l^2} \phi (x \prec^* y) \sum_{k=0}^{n(d)} \bigg(C_{d;k} \sum_{z \prec^*w \in L_{\Lambda,k} (x \prec^* y)} \phi (z \prec^* w) \bigg) \label{EdgeLagrangian} \end{equation} 
Since each point is statistically expected to take up $d$-volume of $l^d$, the pair of points is statistically expected to take the $2d$-volume $l^{2d}$. Therefore, the action is 
\begin{equation} S_{\Lambda} (\phi) = l^{2d} \sum_{x \prec^* y} {\cal L}_{\Lambda} (\phi; x \prec^* y) \end{equation} 
Let us define the set of pairs of edges, $\pi_k$, as
\begin{equation} \pi_{\Lambda, k} = \bigg\{(z \prec^* w, x \prec^* y) \bigg\vert \sharp I_{\Lambda} \Big((z \prec^* w), (x \prec^* y) \Big) = k+1 \bigg\} \end{equation}
If we recall that 
\begin{equation} L_{\Lambda, k} (x \prec^* y) = \bigg\{ (z \prec^* w) \bigg\vert \sharp I_{\Lambda} \Big((z \prec^* w), (x \prec^* y) \Big) = k+1 \bigg\} \end{equation}
it is clear that 
\begin{equation} (z \prec^* w, x \prec^* y) \in \pi_{\Lambda, k} \Longleftrightarrow z \prec^* w \in  L_{\Lambda,k} (x \prec^* y) \end{equation}
and, therefore, we can write the action as
\begin{equation} S_{\Lambda} (\phi) = - \frac{m^2}{2} \sum_{x \prec^* y} \phi^2 (x \prec^* y) - \frac{\lambda}{4!} \sum_{x \prec^* y} \phi^4 (x \prec^* y) - \nonumber \end{equation}
\begin{equation} - \frac{1}{2l^2} \sum_{k=0}^{n(d)} \sum_{((x \prec^*y), (z \prec^* w)) \in \pi_{\Lambda,k}} C_{d;k} \phi (x \prec^* y) \phi (z  \prec^* w) \label{EdgeAction} \end{equation} 
Similarly, in the equation for the weight function we replace the integral with the sum to obtain
\begin{equation} w(\phi, \phi_{ob}) = - \frac{\alpha l^{2d}}{2} \sum_{x \prec^* y} \big(\phi (x \prec^* y) - \phi_{ob} (x \prec^* y) \big)^2 \end{equation} 
We then define the volume element on the function space to be 
\begin{equation} [{\cal D} \phi] = \prod_{x \prec^* y} d \phi (x \prec^* y) \end{equation} 
and write 
\begin{equation} Z_{\Lambda} (\phi_{ob}) = \int [{\cal D} \phi] w(\phi, \phi_{ob}) e^{iS_{\Lambda}(\phi)} \end{equation} 
and then postulate Born's rule to be 
\begin{equation} \rho_{\Lambda} (\phi_{ob}) = \frac{\vert Z_{\Lambda} (\phi_{ob}) \vert^2}{\int[{\cal D} \phi_{ob}] \vert Z_{\Lambda} (\phi_{ob}) \vert^2} \label{BornBoundedCausal}\end{equation}
We will be viewing $\Lambda$ as a physical constant, that happens to be integer-valued. This constant determines the yes-or-no answer of what edge is coupled to what. Since that answer is crucial both for the partition function $Z$, probability density $\rho$, as well as for the quantum state $\psi$ (discussed later), we will be writing $Z_{\Lambda}$, $\rho_{\Lambda}$ and $\psi_{\Lambda}$. Although $\Lambda$ will be dropped during more general discussion. 

\subsection{Removal of the boundary} \label{Conditional}

In case of the absence of the boundary, the Eq \ref{BornBoundedCausal}, as it stands, is divergent. Even if the Lagrangian density converges, the action would diverge due to the integration over spacetime. The conventional approach to this is to assume that the fields attenuate at infinity. But, as we explained in the Introduction, we would like to avoid having to do that. Therefore, we will instead utilize limiting process. In order for the definition of limit to be consistent, we need to introduce two more axioms that causal set needs to obey \cite{Robb1, Robb2, Robb3}:
\begin{equation} \forall r \in {\mathcal C} \forall s \in {\mathcal C} \exists p \in \mathcal{C} ((p \prec r) \wedge (p \prec s)) \label{Robb1} \end{equation} 
\begin{equation} \forall r \in {\mathcal C} \forall s \in {\mathcal C} \exists q \in \mathcal{C} ((r \prec q) \wedge (s \prec q)) \label{Robb2} \end{equation} 
where $\wedge$ means ``and".  We define the limits on a causal set by replacing the ordering $<$ on real line with the partial ordering $\prec$ on a causal set:
\begin{equation} \lim_{p \rightarrow - \infty} f(p) = A \Longleftrightarrow \forall \delta > 0 \exists P \in {\mathcal C} \forall p \prec P (\vert f(p)-A \vert < \delta) \end{equation}
\begin{equation} \lim_{q \rightarrow \infty} f(q) = B \Longleftrightarrow \forall \delta>0 \exists Q \in {\mathcal C} \forall q \succ Q (\vert f(q) - B \vert < \delta) \end{equation} 
\begin{equation} \lim_{p \rightarrow - \infty, q \rightarrow \infty} f(p,q) = C \Longleftrightarrow \forall \delta >0 \exists P \in {\mathcal C} \exists Q \in {\mathcal C} \forall p \prec P \forall q \succ Q (\vert f(p,q) - C \vert < \delta) \end{equation} 
\begin{equation} \limsup_{p \rightarrow - \infty} f(p) = A \Longleftrightarrow \forall \delta > 0 \exists P \in {\mathcal C} \forall p \prec P (\vert \sup \{f(r) \vert r \prec p \} -A \vert < \delta) \end{equation}
\begin{equation} \liminf_{p \rightarrow - \infty} f(p) = A \Longleftrightarrow \forall \delta > 0 \exists P \in {\mathcal C} \forall p \prec P (\vert \inf \{f(r) \vert r \prec p \} -A \vert < \delta) \end{equation}
\begin{equation} \limsup_{q \rightarrow \infty} f(q) = B \Longleftrightarrow \forall \delta>0 \exists Q \in {\mathcal C} \forall q \succ Q (\vert \sup \{f(r) \vert r \succ q \} - B \vert < \delta) \end{equation} 
\begin{equation} \liminf_{q \rightarrow \infty} f(q) = B \Longleftrightarrow \forall \delta>0 \exists Q \in {\mathcal C} \forall q \succ Q (\vert \inf \{f(r) \vert r \succ q \} - B \vert < \delta) \end{equation} 
\begin{equation} \limsup_{p \rightarrow - \infty, q \rightarrow \infty} f(p,q) = C \Longleftrightarrow \nonumber \end{equation}
\begin{equation} \Longleftrightarrow \forall \delta >0 \exists P \in {\mathcal C} \exists Q \in {\mathcal C} \forall p \prec P \forall q \succ Q (\vert \sup \{f(r,s) \vert r \prec p \prec q \prec s \} - C \vert < \delta) \end{equation} 
\begin{equation} \liminf_{p \rightarrow - \infty, q \rightarrow \infty} f(p,q) = C \Longleftrightarrow \nonumber \end{equation}
\begin{equation} \Longleftrightarrow \forall \delta >0 \exists P \in {\mathcal C} \exists Q \in {\mathcal C} \forall p \prec P \forall q \succ Q (\vert \inf \{f(r,s) \vert r \prec p \prec q \prec s \} - C \vert < \delta) \end{equation} 
Thus, heuristically speaking, $- \infty$ is ``infinite past" while $\infty$ is ``infinite future". We used $\delta$ instead of $\epsilon$ because $\epsilon$ has already been used in Section \ref{EpsilonContinuum} with a different meaning. One can show that, as long as the axioms described in Eq \ref{Robb1} and \ref{Robb2} hold, the limits can't be equal to two separate values at the same time. In addition to that, for any given $U \subset \mathcal C$, we will define the \emph{set of edges over U} to be 
\begin{equation} E(U) = \{p \prec^* q \vert p \in U, q \in U \} \end{equation}
We now propose to replace Eq \ref{BornBoundedCausal} with 
\begin{equation} \rho_{\Lambda} (\phi_{ob} (\Omega) \vert \phi_{ob} (E({\cal C}) \setminus \Omega)) = \bigg(\int [{\mathcal D} \phi_{ob} (\Omega)] \Big(\lambda \limsup_{p \rightarrow -\infty, q \rightarrow \infty} \vert Z_{\Lambda} (\phi_{ob} (\Omega \cap E(I(p,q))) \cup \phi_{ob} (E(I(p,q)) \setminus \Omega)) \vert^2 + \nonumber \end{equation}
\begin{equation} +  (1-\lambda) \liminf_{p \rightarrow - \infty, q \rightarrow \infty} \vert Z_{\Lambda} (\phi_{ob} (\Omega \cap E(I(p,q))) \cup \phi_{ob} (E(I(p,q)) \setminus \Omega)) \vert^2 \Big) \bigg)^{-1} \times \nonumber \end{equation}
\begin{equation} \times \Big( \lambda \limsup_{p \rightarrow - \infty, q \rightarrow \infty} \vert Z_{\Lambda} (\phi_{ob} (\Omega \cap E(I(p,q))) \cup \phi_{ob} (E(I(p,q)) \setminus \Omega)) \vert^2 + \nonumber \end{equation}
\begin{equation} + (1-\lambda) \liminf_{p \rightarrow - \infty, q \rightarrow \infty} \vert Z_{\Lambda} (\phi_{ob} (\Omega \cap E(I(p,q))) \cup \phi_{ob} (E(I(p,q)) \setminus \Omega)) \vert^2 \Big) \label{LimSupLimInf} \end{equation}
where $\rho_{\Lambda} (a \vert b)$ denotes the conditional probability of ``$a$" under the assumption ``$b$", given the specific value of the parameter $\Lambda$. The coefficient $\lambda \in [0,1]$ is some agreed-upon constant that we are going to discuss shortly. 

From the damping in Eq \ref{ContinuumAttenuation}, we have physical reasons to expect that $\limsup$ and $\liminf$ would coincide. If such is the case, then the linear combination of $\limsup$ and $\liminf$ would be replaced with a single limit, and the choice of $\lambda$ would be irrelevant. However, proving this mathematically is extremely difficult. After all, Eq \ref{ContinuumAttenuation} was derived in the continuum case, yet we are now talking about discrete case. Intuitively we know that if a discrete set is produced through Poisson process on the continuum, then its various properties would approximate various properties in a continuum with probability that would approach $1$ in appropriate limits. So, intuitively, we might expect that $\limsup$ and $\liminf$ would coincide with a probability $1$. But this intuitive claim would need a mathematical proof and, as it stands, we do not know how to either prove it or disprove it. Therefore, for the purposes of this paper, we will stick with Eq \ref{LimSupLimInf}, and we will leave the above-stated question for the future research. 

If we do stick to Eq \ref{LimSupLimInf}, then we need to discuss various choices of $\lambda$. One argument in favor of $\lambda=\frac{1}{2}$ is an observation that it would allow us to swap limit and integral signs in the following sense. On the one hand, regardless of the choice of $\lambda$, we have 
\begin{equation} \int [{\cal D} \phi_{ob} (\Omega)] \rho_{\Lambda} (\phi_{ob} (\Omega) \vert \phi_{ob} (E({\cal C}) \setminus \Omega)) = 1 \label{Normalization} \end{equation}
where $\rho$ is given by Eq \ref{LimSupLimInf}. But, on other other hand, specifically for $\lambda= \frac{1}{2}$, we have 
\begin{equation} \int [{\mathcal D} \phi_{ob} (\Omega)] \bigg( \frac{1}{2} \limsup_{\Omega \subset E \rightarrow \infty} \frac{\vert Z_{\Lambda} (\phi_{ob} (\Omega) \cup \phi_{ob} (E ({\mathcal C}) \setminus \Omega)) \vert^2}{\int d \phi_{ob} (\Omega) \vert Z_{\Lambda} (\phi_{ob} (\Omega) \cup \phi_{ob} (E ({\mathcal C}) \setminus \Omega)) \vert^2} + \nonumber \end{equation}
\begin{equation} + \frac{1}{2} \liminf_{\Omega \subset E \rightarrow \infty} \frac{\vert Z_{\Lambda} (\phi_{ob} (\Omega) \cup \phi_{ob} (E({\mathcal C}) \setminus \Omega)) \vert^2}{\int d \phi_{ob} (\Omega) \vert Z_{\Lambda} (\phi_{ob} (\Omega) \cup \phi_{ob} (E ({\mathcal C}) \setminus \Omega)) \vert^2} \bigg) = 1 \label{LimSupLimInfHalf}\end{equation} 
Let us explain why that is the case. Suppose $\xi$ denotes an element of some partially ordered set, and suppose there are two functions $\xi \mapsto a_{\xi}$ and $\xi \mapsto b_{\xi}$, such that $a_{\xi} + b_{\xi} = c$ for all $\xi$. Suppose $\limsup a_{\xi} =A$, $\liminf a_{\xi} =a$, $\limsup b_{\xi} = B$ and $\liminf b_{\xi} = b$. Consider the sequence $\xi_n$ such that $a_{\xi_n} \rightarrow A$. Since $\liminf b_{\xi_n} \geq \liminf b_{\xi} = b$, we have $\liminf (a_{\xi_n} + b_{\xi_n}) \geq A + b$. Since $a_{\xi_n} + b_{\xi_n}=c$, we know that $\liminf (a_{\xi_n} + b_{\xi_n}) =c$ and, therefore, $c \geq A+b$. Now consider the sequence $\eta_n$ such that $b_{\eta_n} \rightarrow b$. Since $\limsup a_{\eta_n} \leq \limsup a_{\eta} = A$ we have $\limsup (a_{\eta_n} + b_{\eta_n}) \leq  A+b$. Since $a_{\eta_n}+b_{\eta_n}=c$ we know that $\limsup (a_{\eta_n} + b_{\eta_n}) =c$ and, therefore, $c \leq A+b$. But we said earlier $c \geq A+b$. So we conclude $c=A+b$. By swapping around $a$-s and $b$-s, we can also show that $c=B+a$. Therefore, $2c= (A+b)+(B+a)= (A+a)+(B+b)$ which implies that $c= \frac{A+a}{2} + \frac{B+b}{2}$. Now, we can generalize this from two terms to $N$ terms. Suppose we have $N$ different sequences $\xi \mapsto a_{k \xi}$ for each $k \in \{1, \cdots, N \}$ and suppose that, for each $\xi$, $a_{1 \xi} + \cdots + a_{N \xi} = c$. By using the result for the sum of two terms we can show, by induction, that $c= \frac{A_1+a_1}{2} + \cdots + \frac{A_N+a_n}{2}$, where $A_k = \limsup_{\xi \rightarrow \infty} a_{k \xi}$ and $a_k = \liminf_{k \rightarrow \infty} a_{k \xi}$. Finally, if we replace the sum with an integral, replace $k$ with $\eta$ and repalce $c$ with $1$, we can show that if $\int \rho_{\xi} (\eta) d \eta = 1$ then $\int (\frac{1}{2} \limsup_{\xi \rightarrow \infty} \rho_{\xi} (\eta) + \frac{1}{2} \liminf_{\xi \rightarrow \infty} \rho_{\xi} (\eta)) d \eta = 1$. From this, we can convince ourselves of Eq \ref{LimSupLimInfHalf} by simply noticing that the integral of the common expression under $\limsup$ and $\liminf$ is  $1$.  

To summarize what we just said, our prime motive of choosing $\lambda = \frac{1}{2}$ is our ability to swap limits and integration signs and obtain Eq \ref{LimSupLimInfHalf}. However, this feature is not necessary. In case of $\lambda \neq \frac{1}{2}$, Eq \ref{LimSupLimInf} will still be well defined and Eq \ref{Normalization} will still hold, which is all we need. This being the case, let us discuss another choice of $\lambda$; namely, $\lambda=1$. While there is no mathematical reason to make that choice, there is a hand-waving motivation in its favor. We can think of $I (p,q) \supset \Omega$ as an ``explanation" of how a specific trajectory $\phi (\Omega)$ was produced. The nature of scientific method is that we are looking for ``best possible explanation" and throwing away all the other ones. As it is, there is no ``best possible explanation" but there is a sequence of ``better and better explanations" corresponding to the process of taking $\limsup$ -- and $\limsup$ is what we obtain by taking $\lambda=1$. Heuristically speaking, we are looking at all possible $\phi (\Omega)$; for each one we ask ourselves ``what is the most likely scenario that would produce it" and then we compare the probabilities of these ``most likely scenarios". This explanation, as it stands, would not stand to scrutiny. For one thing, since there are several different $I(p,q)$ that can potentially create $\phi (\Omega)$, one has to find a way of ``comparing their numbers" (which are infinite). One heuristic way of dealing with this is to claim that that number is actually $1$ -- in particular, we have one single Alexandrov set with infinite size that is being approximated by an infinite sequence of Alexandrov sets of finite size (such techniques are widely used in nonstandard analysis \cite{Nonstandard1, Nonstandard2}). This, however, is still very much hand waving. But that is okay since we are not claiming to make any mathematical statements here, we are only suggesting a motivation for the choice of $\lambda=1$. From strictly mathematical point of view, we can simply choose $\lambda=1$ without any reason given. But, as an optional part, we can provide hand-waving motivations for it, that can be as hand wavy as we like. 

\section{Quantum States} \label{QuantumStates}

\subsection{Minimalist model of quantum states when $\mathcal C$ is bounded} \label{StatesBoundedMinimalist}

In Section \ref{Conditional} we were looking at the region $\Omega \subset E({\mathcal C})$, where $\Omega$ is bounded and $\mathcal C$ is unbounded. We were addressing the fact that $\mathcal C$ is unbounded by using the limitting process. However, there is an alternative approach. We can assume unbounded $\mathcal C$, single out the bounded region $\Omega \subset E({\mathcal C})$, defined the thickened boundary $\Sigma$ of $\Omega$ and assign quantum state to $\Sigma$. That quantum state will include all the relevant information in $E({\mathcal C}) \setminus \Omega$ that is relevant to $\Omega \setminus \Sigma$. The word ``relevant" is the key word. Since $\mathcal C$ is infinite while $\Sigma$ is finite, $\Sigma$ can not possibly contain all the information contained in $\mathcal C$. But, as it turns out, only finite amount of that infinite information is relevant to $\Omega \setminus \Sigma$, and quantum state assigned to $\Sigma$ \emph{can} contain that, \emph{finite}, information. As we just indicated, our ultimate goal is to do that when $\mathcal C$ is unbounded.  However, since dealing with unbounded $\mathcal C$ creates some extra concerns that need to be addressed, let us focus in the current section on the toy model when $\mathcal C$ is bounded -- and then we will discuss the situations with unbounded $\mathcal C$ in the Sections \ref{StatesUnboundedMinimalist}-\ref{Causality}. 

If one inspects Eq \ref{EdgeDambertan} one can see that some of the terms in the summation represent couplings between edges that are up to $n(d)$ distance apart from each other. So if the thickness of $\Sigma$ is smaller than $n(d)$, some parts of the signal will be able to ``jump" across $\Sigma$ without affecting the state attached to $\Sigma$. To be sure, there are other terms that ``jump" by the smaller distance, or none at all. So the state attached to $\Sigma$ will be affected \emph{somehow}. But due to missing \emph{some} parts of the signal, it will not contain all the relevant information we might need. In order to avoid this situation, we will assume that the thickness of $\Sigma$ is at least $n(d)+1$. This will require for us to define a ``thickened boundary". We will define it as follows

{\bf Definition:} Suppose $\Sigma \subseteq \Omega \subset E({\mathcal C})$. We say that $\Sigma$ is a boundary of $\Omega$ that is \emph{thicker than $N$} if one can \emph{not} find $z \prec^* w \in L_{N+1} (x \prec^* y)$ such that \emph{either} (i) $z \prec^* w \in E({\mathcal C}) \setminus \Omega$ and $x \prec^*y \in \Omega \setminus \Sigma$ \emph{or} (ii) $z \prec^*w \in \Omega \setminus \Sigma$ and $x \prec^* w \in E({\mathcal C}) \setminus \Omega$. 

{\bf Definition} Suppose $\Sigma \subseteq \Omega \subset E({\mathcal C})$. We say that $\Sigma$ is a boundary of $\Omega$ \emph{of thickness $N$} if the following two statements are true:

a) $\Sigma$ is a boundary of $\Omega$ that is thicker than $N-1$

b) For any $(p \prec^* q) \in \Sigma$, the set $\Sigma \setminus \{p \prec^* q \}$ is \emph{not} a boundary of $\Omega$ of thickness greater than $N-1$. 

In other words, $\Sigma$ is a \emph{minimal} set that can be described as a boundary of $\Omega$ thicker than $N-1$. 

Let us assume from this point onward that $\Sigma$ is a boundary of $\Omega$ of thickness $n(d)+1$. We can then split $\pi_{\Lambda,k}$ as follows:
\begin{equation} \pi_{\Lambda,k} = \pi_{\Lambda,k}^{in} (\Omega) \cup \pi_{\Lambda,k}^{out} (\Omega) \end{equation}
where 
\begin{equation} \pi_{\Lambda,k}^{in} (\Omega) = \pi_{\Lambda,k} \cap (\Omega \times \Omega) \end{equation}
\begin{equation} \pi_{\Lambda,k}^{out} (\Omega) = \pi_{\Lambda,k} \setminus (\Omega \times \Omega) \end{equation}
and express Eq \ref{EdgeAction} as 
\begin{equation} S_{\Lambda} (\phi) = S_{\Lambda,in} (\phi, \Omega) + S_{\Lambda,out} (\phi, \Omega) \end{equation}
where
\begin{equation} S_{\Lambda,in} (\phi, \Omega) = - \frac{m^2}{2} \sum_{x \prec^* y \in \Omega} \phi^2 (x \prec^* y) - \frac{\lambda}{4!} \sum_{x \prec^* y \in \Omega} \phi^4 (x \prec^* y) - \nonumber \end{equation}
\begin{equation} - \frac{1}{2l^2} \sum_{k=0}^{n(d)} \sum_{((x \prec^*y), (z \prec^* w)) \in \pi_{\Lambda,k}^{in}} C_{d;k} \phi (x \prec^* y) \phi (z  \prec^* w) \end{equation} 
\begin{equation} S_{\Lambda,out} (\phi, \Omega) = - \frac{m^2}{2} \sum_{x \prec^* y \in E({\mathcal C}) \setminus \Omega} \phi^2 (x \prec^* y) - \frac{\lambda}{4!} \sum_{x \prec^* y \in E({\mathcal C}) \setminus \Omega} \phi^4 (x \prec^* y) - \nonumber \end{equation}
\begin{equation} - \frac{1}{2l^2} \sum_{k=0}^{n(d)} \sum_{((x \prec^*y), (z \prec^* w)) \in \pi_{\Lambda,k}^{out}} C_{d;k} \phi (x \prec^* y) \phi (z  \prec^* w) \end{equation} 
where the source of $\Lambda$ in $S_{\Lambda}$ on the left hand side is the occurence of $\Lambda$ in $\pi_{\Lambda;k}$ under one of the summation signs on the right hand side. In light of the fact that the thickness of $\Sigma$ is $n(d)+1$, $S_{\Lambda,out} (\phi, \Omega)$ is a function of $\phi (\Sigma) \cup \phi (E(C) \setminus \Omega)$ alone: 
\begin{equation} S_{\Lambda,out} (\phi (E({\mathcal C}), \Omega)) = S_{\Lambda,out} (\phi (\Sigma) \cup \phi (E(C) \setminus \Omega)) \end{equation}
In light of this, we can write 
\begin{equation} \int [{\mathcal D} \phi (E({\mathcal C}))] w(\phi (E({\mathcal C})), \phi_{ob} (E({\mathcal C}))) \exp (iS_{\Lambda}(\phi(E({\mathcal C})))) = \nonumber \end{equation}
\begin{equation} = \int [{\mathcal D} \phi (\Omega)] \psi_{\Lambda} (\phi (\Sigma))  w(\phi (\Omega), \phi_{ob} (\Omega)) \exp (iS_{\Lambda, in} (\phi(\Omega))) \end{equation}
where 
\begin{equation} \psi_{\Lambda} (\phi (\Sigma)) = \int [{\mathcal D} \phi (E({\mathcal C}) \setminus \Omega)] w ( \phi (E({\mathcal C}) \setminus \Omega),  \phi_{ob} (E({\mathcal C}) \setminus \Omega)) \exp (iS_{\Lambda,out} (\phi,\Omega)) \label{CtoPsi} \end{equation}
We claim that $\psi_{\Lambda} (\phi (\Sigma))$ can be viewed as a \emph{quantum state} attached to $\Sigma$. Let us motivate this viewpoint. Recall that, in case of first quantization, quantum state is identified with a wave function $\psi (x)$. In case of second quantization, the quantization of $x$ is being replaced by the quantization of $\phi$. Therefore, $\psi (x)$ should be replaced with $\psi (\phi)$. However, we should be careful. A four-dimensional trajectory of $\phi$ corresponds to a \emph{trajectory} $x(t)$ as opposed to a single point $x$. The analogue of a single point $x$ is the \emph{slice} of $\phi$ on a \emph{hypersurface} $\Sigma$, that is, $\phi (\Sigma)$. Therefore, $\psi (x)$ should be replaced with $\psi_{\Lambda} (\phi (\Sigma))$. That is to be interpreted as a quantum state attached to $\Sigma$. As mentioned earlier, $\Lambda$ is a constant, so by $\psi_{\Lambda} (\phi (\Sigma))$ we mean one single state as opposed to a family of states. 

\subsection{Quantum states when $\mathcal C$ is unbounded} \label{StatesUnboundedMinimalist}

Let us now attempt to generalize what we did in Section \ref{StatesBoundedMinimalist} to the situation when the quantum states are unbounded. We would like to employ similar limitting process as we did in Section \ref{Conditional}. However, as we recall from Section \ref{StatesBoundedMinimalist}, we were unable to prove the existence of the limit so we had to replace it with a linear combination of $\limsup$ and $\liminf$. This translates into us being unable to describe it as a single state in a sense of Section \ref{QuantumStates}. Instead, we will introduce the notion of a \emph{hyperstate}. We will introduce it in the following way:

 {\bf Definition} Let $({\cal H},d)$ be a normed Hilbert space, and let $\Pi$ be partially ordered set. A \emph{hyperstate extension} of $({\cal H},d)$ through $\Pi$ is another normed Hilbert space, $({\cal H}_{\Pi}, d_{\Pi})$, where ${\cal H}_{\Pi}$ is a set of maps of the form $\Pi \mapsto \cal H$ and $d_{\Pi}$ is defined as $d_{\Pi} (u,v) = \limsup_{\xi \rightarrow \infty} d(u(\xi),v(\xi))$. Furthermore, there is a natural embedding of $\cal H$ into ${\cal H}_{\Pi}$: for any $h \in \cal H$, we define $\overline{h} \in {\cal H}_{\Pi}$ as $\overline{h} (\xi) = h$ for all $\xi \in \Pi$. 

Now, we will define ${\cal H}_{\Sigma}$ to be the set of states over $\Sigma$ with the addition, multiplication and norm defined in the usual way
\begin{equation} (\psi_1 + \psi_2)(\phi (\Sigma)) = \psi_1 (\phi (\Sigma)) + \psi_2 (\phi (\Sigma)) \end{equation}
\begin{equation} (c \psi) (\phi (\Sigma)) = c( \psi (\phi (\Sigma))) \end{equation}
\begin{equation} \vert \psi \vert = \bigg( \int [{\mathcal D} \phi (\Sigma)] (\psi (\phi))^2 \bigg)^{1/2} \end{equation}
and we will define $\Pi_{\Sigma}$ to be the set of intervals containing $\Sigma$, ordered by inclusion:
\begin{equation} \Pi_{\Sigma} = \{ I (p,q) \supset \Sigma \} \end{equation}
\begin{equation} I (p_1, q_1) \leq I(p_2,q_2) \Longleftrightarrow I(p_1,q_1) \subseteq I(p_2,q_2) \end{equation} 
Now, we will define a hyperstate $\Psi$ as follows. For any $I(p,q)$ we define $\psi = \Psi(I(p,q))$ by Eq \ref{CtoPsi} with $C$ being replaced with $I(p,q)$:
\begin{equation} (\Psi_{\Lambda} (I(p,q))) (\phi (\Sigma)) = \end{equation}
\begin{equation} = \int [{\mathcal D} \phi (E(I(p,q)) \setminus \Omega) w ( \phi (E(I(p,q)) \setminus \Omega),  \phi_{ob} (E(I(p,q)) \setminus \Omega)) \exp (iS_{\Lambda} (\phi (E(I(p,q)) \setminus \Omega))) \nonumber \end{equation} 
On the first glance it seems troublesome. Our original goal was to ``reduce" the infinite information contained in $\cal C$ to the finite information contained in $\psi$. We have not accomplished that goal: the $\Psi$ that is defined above contains \emph{infinite} information, since it incorporates infinitely many $I(p,q)$. However, thanks to the definition of the distance, we \emph{do} have some measns of going around this objection. In particular, while $\Psi$ contains infinite information, we could say that $\Psi$ can be \emph{approximated} by $\psi$, where $\psi$ contains finite information. Here, by ``approximation" we mean $d(\Psi, \overline{\psi}) < \delta$, for some suitable $\delta$. Of course, this does not need to be the case. But, in light of the damping in Eq \ref{ContinuumAttenuation}, we would expect on the physical groups that the fluctuations of $\Psi (I(p,q))$ will be small once $I (p,q)$ is large enough. In fact, we would expect those fluctuations to approach zero; but since we are unable to prove this (which is the reason why we had to resort to $\limsup$ and $\liminf$ to begin with) we have to instead stick to the ``middle ground" and just assume they are small.

We will now carry over Eq \ref{LimSupLimInf} into our framework. First of all, we will replace $Z_{\Lambda}(\phi_{ob} (E(I(p,q))))$ with $Z_{\Lambda}(\Psi (\Sigma), \phi_{ob} (E(\Omega)))$. That is, we replace $\phi (I(p,q))$ with $\phi (\Omega)$ (where $\Omega \subset I (p,q)$) and take care of the ``missing information" by including $\Psi (\Sigma)$. We define $Z_{\Lambda}(\Psi (\Sigma), \phi_{ob} (E(\Omega))$ to be 
\begin{equation} Z_{\Lambda}(\Psi (\Sigma), \phi_{ob} (\Omega)) = \int [{\cal D} \phi (\Omega)] \Psi (\phi (\Sigma)) w(\phi_{ob} (\Omega), \phi (\Omega)) \exp (iS (\phi (\Omega))) \end{equation}
In addition to $d_{\Pi}$ defined earlier, we also define $d^*_{\Pi}$ as $d^*_{\Pi} (u,v) = \liminf_{\xi \rightarrow \infty} d(u(\xi),v(\xi))$. Note that while $d_{\Pi}$ obeys the triangle inequality, $d^*_{\Pi}$ does not. Thus, $d_{\Pi}$ is a distance function, but $d^*_{\Pi}$ is \emph{not}. We can then rewrite Eq \ref{LimSupLimInf} as 
\begin{equation} \rho_{\Lambda} (\phi_{ob} (\Omega) \vert \phi_{ob} (E({\cal C}) \setminus \Omega)) = \nonumber \end{equation}
\begin{equation} = \bigg(\int [{\mathcal D} \phi_{ob} (\Omega)] \Big(\lambda d^2_{\Pi} (0, Z_{\Lambda} (\Psi (\Sigma), \phi_{ob} (\Omega)))  + (1- \lambda) d^{*2}_{\Pi} (0, Z_{\Lambda} (\Psi (\Sigma), \phi_{ob} (\Omega))) \Big) \bigg)^{-1} \times  \nonumber \end{equation}
\begin{equation} \times  \Big(\lambda d^2_{\Pi} (0, Z_{\Lambda} (\Psi (\Sigma), \phi_{ob} (\Omega)))  + (1- \lambda) d^{*2}_{\Pi} (0, Z_{\Lambda} (\Psi (\Sigma), \phi_{ob} (\Omega))) \Big) \label{Exact} \end{equation} 
If it happens that $d_{\Sigma} (\Psi, \overline{\psi}) < \delta$ for some state $\psi (\phi (\Sigma))$, then we can obtain a good approximation to the above equation by replacing $\Psi$ with $\overline{\psi}$. In this case, it reduces to 
\begin{equation} \rho_{\Lambda} (\phi_{ob} (\Omega) \vert \phi_{ob} (E({\cal C}) \setminus \Omega)) \approx \frac{\vert Z_{\Lambda}(\Psi (\Sigma), \phi_{ob} (\Omega)) \vert^2 }{\int [{\cal D} \phi_{ob} (\Omega)]\vert Z_{\Lambda}(\Psi (\Sigma), \phi_{ob} (\Omega)) \vert^2 } \label{Approx} \end{equation} 
While Eq \ref{Exact} is exact, Eq \ref{Approx} is only an approximation. But, at the same time, Eq \ref{Exact} utilizes infinite information while Eq \ref{Approx} utilizes finite information. Thus, we have just formulated more precisely what we intuitively knew all along: while the exact calculation should take into account an entanglement across infinite universe, it can be approximated by a calculation done within a finite region of that universe. For example, for most everyday purposes, we can pretend that the universe is just our galaxy, although there might be some really small errors due to our neglecting the signals coming from the outside of our galaxy. 

\subsection{More general model of quantum states} \label{StatesGeneral}

In Sections \ref{StatesBoundedMinimalist} and \ref{StatesUnboundedMinimalist} we have assumed that the quantum states attached to $\Sigma$ were produced by $\phi_{ob}$ in in the exterior of $\Sigma$. Since that exterior is infinite, this forced us to propose the ways of handling such infinity which resulted in us having to introduce extra formalism specifically for that purpose. There is an alternative approach, however. We can drop the assumption that the quantum states on $\Sigma$ were created by $\phi_{ob}$ in the exterior and, instead, simply postulate those states. The role of $\phi_{ob}$ is to influence the way $\psi (\phi)$ \emph{evolves} from one hypersurface to the next, whereas $\psi (\phi)$ at one \emph{specific} hypersurface can be thought of as ``initial condition". Since we are free to postulate any initial conditions we like, we are free to introduce $\psi (\phi)$ as a \emph{regular} state, \emph{not} a hyperstate. 

Suppose $\Omega_1 \subseteq \Omega_2$ and suppose $\Sigma_1$ and $\Sigma_2$ are boundaries of thickness $n(d)$ of $\Omega_1$ and $\Omega_2$, respectively. Furthermore, suppose that $\Omega_1 \setminus \Sigma_1 \supseteq \Omega_2 \setminus \Sigma_2$. We postulate the following relation: 
\begin{equation} \psi (\phi (\Sigma)) = \int [{\mathcal D} \phi (\Omega_1 \setminus \Omega_2)] \psi (\phi (\Sigma_1)) w ( \phi (\Omega_1 \setminus \Omega_2),  \phi_{ob} (\Omega_1 \setminus \Omega_2)) \exp (iS_{\Lambda,out} (\phi,\Omega_2))  \end{equation}
The reason we write $\psi$ rather than $\psi_{\Lambda}$ is that the source of $\Lambda$ in $\psi_{\Lambda}$ used to be the action at $E({\cal C}) \setminus \Omega$, and that action was based on $\Lambda$. But with what we are doing now, that action is no longer relevant, since we are merely postulating $\psi$. That is why $\psi$ no longer has $\Lambda$ index in it. However, the \emph{relation} between $\psi$-s on two \emph{different} hypersurfaces still depends on the action \emph{between} these hypersurfaces, which, in turn, depends on $\Lambda$. That is why the above equation still has $\Lambda$ in it. 

What we have done in Sections \ref{StatesBoundedMinimalist} and \ref{StatesUnboundedMinimalist} can be heuristically viewed as a ``special case" of the current section for the situation where $\psi (\phi (\Sigma)) \rightarrow 1$ as $\Sigma \rightarrow \infty$ (where by $\Sigma \rightarrow \infty$ we mean that $\Omega_k \setminus \Sigma_k \supset I (r_k, s_k)$ where $r_k \rightarrow - \infty$ and $s_k \rightarrow \infty$ as $k \rightarrow \infty$). At the same time, however, it seems appealing to be able to stick to Sections  \ref{StatesBoundedMinimalist} and \ref{StatesUnboundedMinimalist} without having to resort to the current section. For one thing, one can visualize $\phi_{ob}$ while one can not visualize $\psi (\phi)$, so it would be nice to be able to say that the latter reduces to the former. So we will devote the next section to discussing pros and cons in choosing between Sections  \ref{StatesBoundedMinimalist} and \ref{StatesUnboundedMinimalist} and the current section. 

\subsection{Choice between Section \ref{StatesGeneral} versus Sections \ref{StatesBoundedMinimalist} and \ref{StatesUnboundedMinimalist}} \label{Choice}

We have just said that the model described in Sections \ref{StatesGeneral} to Sections \ref{StatesBoundedMinimalist} is a special case of the model described in Section \ref{StatesGeneral}. At the same time, however, one can argue that all of the observed phenomena can be described within the framework of that ``special case". After all, nobody has ever seen $\psi (\phi)$; what people observe is $\phi_{ob}$. For example, an arrow of measuring apparatus pointing in a certain direction can be described as $\phi_{ob}$ being larger within a region of the shape of that arrow and smaller outside of that region. Based on the behavior of $\phi_{ob}$ we then \emph{infer} what $\psi (\phi)$ might be. But, as it turns out, this inference is not one to one: one would have to ``know" the behavior of $\psi (\phi)$ at infinity in order to establish that correspondence. That is what justifies us making a default assumption about it, such as it is a constant. On the other hand, one can also argue that even if $\psi (\phi)$ at infnity was not a constant, its effects would have decayed due to the damping in Eq \ref{ContinuumAttenuation}. In this case, one does \emph{not} need to make any assumption about $\psi (\phi)$ at infinity, although making one is not going to alter the result. In other words, we know that that assumption won't cause any trouble; the only question is whether or not it is necessary. That question is tied to the question whether the limit discussed earlier is well defined (as a consequence of the damping in Eq \ref{ContinuumAttenuation}) or whether we should resort to $\limsup$ and $\liminf$. On the physical grounds, we would like to say that limit is well defined and also that the situation is independent of $\psi (\phi)$ at infinity. But, since we can not prove it mathematically as of yet, we would have to assume otherwise. 

It should be pointed out that the proposal in this section is more closely alligned to GRW model \cite{GRW1, GRW2, GRWSverdlov} than the proposals of  Sections \ref{StatesBoundedMinimalist} and \ref{StatesUnboundedMinimalist}. In the case of non-relativistic quantum mechanics, GRW model can be described as follows. A wave function $\psi (x)$ evolves according to Schrodinger's equation and, at random times, it is being multiplied by Gaussians, $\psi (x) \mapsto N (x_0, \psi) \psi (x) e^{- \frac{\alpha}{2} (x-x_0)^2}$ where $N (x_0, \psi)$ is a normalization constant that is determined by keeping it norm-1. That multiplication is called the \emph{hit}. The point $x_0$ is chosen randomly; the probability density that the point $x_0$ will fall in the vicinity of $y_0$ is proportional to $\frac{1}{N^2 (y_0, \psi)}$. The repeated multiplication by these Gaussians is responsible for the collapse of wave function. If we have repeated such events, we can replace $x_0$ with $\cdots, x_{-1}, x_0, x_1, \cdots$ to avoid confusion: $x_k$ is the choice of the center of the Gaussian during the event number $k$. The timing of these hits play important role too, since affects the amount by which wave function evolved between the hits. We will denote the time of the hit around $x_k$ to be $t_k$. Thus, the list of hits can be written as $(\cdots, (x_{-1}, t_{-1}), (x_0, t_0), (x_1, t_1), \cdots )$. It has been shown in \cite{GRWSverdlov} that GRW model would approximate continuous measurement model in the limit that the time between these hits goes to zero (and, accordingly, the effect of each hit goes to zero as well, so that the accumulated effect of multiple hits stays fixed). In this case, the discrete sequence $(\cdots, (x_{-1}, t_{-1}), (x_0,t_0), (x_1,t_1), \cdots)$ can be replaced with a continuous trajectory $x_{ob} (t)$. Now, what we just said pertains to the first quantization. In case of the second quantization, $x_{ob} (t)$ gets replaced with $\phi_{ob} (x,t)$, which is what we have been focusing on throughout the rest of the paper. Therefore, the statement that ``$\psi (\phi)$ can be reduced to $\phi_{ob}$" is an analogue of the statement ``$\psi (x)$ can be reduced to $(\cdots, (x_{-1},t_{-1}), (x_0,t_0), (x_1,t_1), \cdots)$". But, in case of GRW model, it is clearly not true. It is assumed that we already have $\psi (x)$, and those Gaussians merely modify its evolution. Therefore, by this logic, one could argue that we need to make an extra assumption about $\psi (\phi)$   as is done in Section \ref{StatesGeneral} much like in GRW model one has to make an extra assumption about $\psi (x)$. 

However, one can also argue in the opposite direction. One can utilize our other arguments in favor of Sections \ref{StatesBoundedMinimalist} and \ref{StatesUnboundedMinimalist} and claim that, by the same token, $\psi (x)$ should be viewed as a bi-product of $(\cdots, (x_{-1}, t_{-1}), (x_0,t_0), (x_1,t_1), \cdots)$. One argument in this direction is that we have not actually observed $\psi (x)$, and neither do we observe localization of the particle either. What we observe are macroscopic objects. That is even true in double slit experiment. When the particle hits the screen, the dot that we see is clearly large enough for us to see it -- in other words, it is macroscopic in size. The fact that it was created by an electron is merely our theory. One can then point out that the Gaussians involved in GRW model might be macroscopic. After all, the effect of each one of the Gaussians is very small and they only create the collapse after their effects accumulate. This implies that their are very wide which, in turn, suggests the possibility that they might be macroscopic in size. Since the Gaussians are macroscopic and what we observe is macroscopic too, this makes it logical to say that Gaussians is what we observe. One initial objection one might have to this is that the ``macroscopic" object we referred to a bit earlier -- namely, the spot on the screen -- is not the same as the ``macroscopic" object we are talking about now -- namely Gaussians. But it is possible to address this objection by saying that we don't directly observe the spot on the screen but, instead, we observe multiple Gaussians that we then interpret as the spot on the screen. If we do resort to this kind of argument, we could then go back and ask how do we know that we observe the Gaussians rather than observing the mean and standard deviation of the wave function instead? The answer is that we don't know. But, at the same time, it is possible to \emph{speculate} that Gaussians is what we observe -- particularly since in \cite{GRWSverdlov} these Gaussians were linked to $x_{ob}$ which corresponds to the observed location of macroscopic object. If we do stick to this hypothesis, we could argue that $\psi$ is only a bi-product of these Gaussians -- and, therefore, drop the initial conditions on $\psi$. We could also argue that, thanks to these Gaussians, the information about initial conditions at $- \infty$ gets lost anyway. We then can use this way of looking at GRW model as a way of justifying a choice in favor of Sections \ref{StatesBoundedMinimalist} and \ref{StatesUnboundedMinimalist}.

Another argument in favor of Sections \ref{StatesBoundedMinimalist} and \ref{StatesUnboundedMinimalist} and against the Section \ref{StatesGeneral} is the following. While it is straightforward to see how $\psi (\phi (\Sigma))$ evolves when $\Sigma$ ``shrinks", it is a lot harder to solve an ``inverse problem" and find out how it evolves when $\Sigma$ ``expands". The good news, however, is that we do not have to assume that there is any space outside of any given hypersurface. More precisely, we didn't say there is, and we didn't say there isn't. So, as far as we are concerned, it is ``possible" that the state on the hypersurface was generated by the bigger hypersurface, and it is also possible that there is no bigger hypersurface, and the state on the hypersurface simply ``created itself" so to speak. Of course, from the point of view of the goal of our project, we would like to think that there is a bigger hypersurface, but we do not have to prove that there is one.

\subsection{Quantum measurement and causality} \label{Causality} 

One thing the reader might have noticed in the comparison with GRW model, is that this comparison can only work if ``time direction" points from the outside of $\Omega$ to the inside. Indeed, if we have $\Omega_1 \supset \Omega_2$, with their respective boundaries $\Sigma_1$ and $\Sigma_2$, then we can use the quantum state assigned to $\Sigma_1$ to make a prediction of $\phi_{ob}$ in $\Sigma_1 \setminus \Sigma_2$ which, in turn, can be used in order to predict the quantum state assigned to $\Sigma_2$. Per earlier discussion, $\phi_{ob}$ can be thought of as an analogue of ``hits" in GRW model, while the quantum states assigned to $\Sigma_1$ and $\Sigma_2$ can be thought of as an initial state and final state, respectively. 

One can re-parametrize the coordinates in a way that would reflect this. That is, $dx^0$ can point in future timelike direction, or past timelike direction, or spacelike direction, depending on which direction points to the center of $\Omega$. This reparametrization does \emph{not} reflect the causal relations $\prec$. The future in terms of $\prec$ can point both to the center of $\Omega$ and away from it, depending on where we are located, while the future in terms of $x^0$ always points to the center of $\Omega$. If one follows the causal lines of $\prec$, one would enter $\Omega$ from one end and leave it from the other end. If one follows the causal lines according to $x^0$, one would enter $\Omega$, reach its center and stop there. We are already familiar with examples of when $dt$ is spacelike from the black holes. In this case, however, we are \emph{not} dealing with black holes or any other notrivial geometry for that matter (for all we know it might be a flat Minkowski space) but still nothing stops us from reparametrizing the coordinates whatever way we like. 

In order to best reflect the kind of $x^0$ that we experience, consider $\sigma \subset \Sigma$. While $\Sigma$ closes onto itself, $\sigma$ does not. In our case, $\sigma$ is part of the ``earlier" side of $\Sigma$, which is the reason why \emph{our} version of $x^0$ points to the future. One can then define the quantum state attached to $\sigma$ by 
\begin{equation} \Psi_{\Lambda} (\phi (\sigma)) = \int [{\mathcal D} \phi (\Sigma \setminus \sigma)] \Psi_{\Lambda} (\phi (\sigma) \cup \phi (\Sigma \setminus \sigma)) \end{equation}
where by $\phi (X) \cup \phi (Y)$ we mean $\phi (X \cup Y)$. Note that, while we \emph{choose} to select $\sigma$ to be a patch on the ``past" part of $\Sigma$ we didn't have to. If we were to choose $\sigma$ at other locations, then the time direction, relevant to $\sigma$ can point to the future with respect to $\prec$, or to the past with respect to $\prec$ or in spacelike direction with respect to $\prec$. If it points in the spacelike direction with respect to $\prec$, then the physics observed by such an observer would be drastically different from the physics as we know it. But, if it points in the timelike past direction, the physics will be the same due to the time reversal symmetry.

Let us see whether the time reversal symmetry is exact or approximate. The Lagrangian density (Eq \ref{EdgeLagrangian}) seems to suggests its approximate, while the action (Eq \ref{EdgeAction}) seems to suggest it is exact. Let us try to reconcile this seeming contradiction. We observe that the terms that break the time reversal symmetry in Eq \ref{EdgeLagrangian} are the terms of the form $\phi (x \prec^* y) \phi (z \prec^* w)$. This term, on its own, respects time reversal symmetry. The point where it is broken is when we ``decide" whether we want to ``count" this term as a part of Lagrangian density at $x \prec^* y$ or as a part of Lagrangian density of $z \prec^* w$. Since the action involves the sum of both of these Lagrangian densities, it does not matter which one we would include any given term into. It is merely the question of convention. In other words, the action represents actual physics while the Lagrangian density represents our choice on how to group the terms. Therefore, since the action respects the exact time reversal symmetry, our theory respects the exact time reversal symmetry as well. 

As far as Lorentz invariance issues, the choice of reference frame is basically the choice of $\sigma$ and $\Sigma \supset \sigma$. The choice of $\sigma$ will tell us that the gradient of $x^0$ will be normal to that surface, and the choice of $\Sigma \supset \sigma$ will tell us which of the two opposite directions it is pointing at. Since our theory accommodates all possible choices of $\sigma$ and $\Sigma \supset \sigma$, it respects the principle of relativity in this particular sense. 

There is also an alternative interpretation to everything we said in this sub-section so far. Namely, there is no such thing as sequence of measurements. Instead, there is one single ``snapshot" taken over the entire spacetime history. In case of bounded causal set, the probability of that snapshot is given by Eq \ref{BornBoundedCausal}. In case of unbounded causal set, the space of trajectories $\phi_{ob}$ became infinite dimensional which made it difficult to talk about the set of probability densities over that space.  So we had to resort to conditional probability densities instead. The role of $\Omega$ is simply our way of formulating those conditional probabilities, but $\Omega$ does not play any physical role. From the physics point of view, we had one single snapshot, where both $\Omega$ and $E({\mathcal C}) \setminus \Omega$ were taken at the same time. 

One way to reconcile these two interpretations is to picture the following scenario. Suppose few different measurements of few different systems take place at the same time. But suppose that the lab technician is unable to look at all these outcomes at the same time. For example, maybe they have to walk from one laboratory to the other; maybe it would take time for them to record the results with paper and pencil; maybe there are some other similar obstacles. So these delays are not quantum mechanical but they are specific to an observer. The actual quantum measurements were made simultaneously, but the observer looks at them in the sequence. In the same way, the actual measurement of the spacetime history was taken simultaneously as one single snapshot. But an observer is unable to look at the whole snapshot at the same time; they have to first look at one part of it and then the other. Different observers make different choices as to what part they look first and what part they look next. These choices correspond to different choices of time coordinate that these different observers chose to parametrize the spacetime history with. 

Note that we link our proposal to GRW model and, at the same time, claim that we respect Lorentz invariance. Therefore, we can think of our proposal as a relativistic version of GRW model, in some very vague sense. It should be pointed out that there was earlier work by Tumulka that also made that claim \cite{RelativisticGRW}. However, that model was using hyperboloids that the ``hits" were creating. In our proposal, on the other hand, we were using $\Sigma$. Our $\Sigma$ is different from their hyperboloids in the following sense. Their hyperboloids were created by the ``hits" whereas our $\Sigma$ was the result of lack of knowledge of the observer (knowledge in the classical sense). Therefore, in their case, all observers would agree what those hyperboloids are, while in our case they would not agree regarding the choice of $\Sigma$. If we stick with the mindset that hypersurfaces create preferred frames, this would imply that in their proposal all observers would agree on the preferred frame, while in our proposal they wouldn't. In this specific respect, our proposal respects principles of relativity more closely. However, even in their case, there were no \emph{a priori} preferred frame. The preferred frame was \emph{spontaneously created} after the ``hits". 

\section{Conclusion}

In this paper we have accomplished two things. First, we have addressed the question of nonlocality of causal sets and, secondly, we proposed the model of quantum measurement in causal set context. Interestingly, the questions regarding principles of relativity arise in both contexts. In the context of causal sets, one attempts to address the question of microscopic violation of relativity by the discrete structure as well as macroscopic violation of relativity by the boundary. In case of quantum measurement, one has to address grandfather's paradox and other similar issues. As it turns out, this is not coincidence. According to Section \ref{Causality}, the quantum states are attached to the concentric hypersurfaces and the direction of time points inward. Therefore, there is a relation between the preferred frame determined by the shape of the hypersurface (causal set problem) and the preferred frame determined by the quantum measurement (foundation of quantum mechanics problem). Both issues are addressed by making sure that causal set is unbounded and, therefore, the choice of that hypersurface is up to the observer as opposed to intrinsic to the causal set itself. 

In the past, there have been proposals of quantum measurement models by causal set theorists (see, for example, \cite{QuantumLogic1, QuantumLogic2}). These approaches, however, were based on introducing \emph{quantum logic}, where the rules of usual logic (referred to as \emph{classical logic}) no longer hold and are, instead, replaced with the rules of what they call \emph{quantum logic}. By sharp contrast to those approaches, this paper is written strictly within the realm of \emph{classical logic}. In addition to this, the notion of quantum states was shown to be emergent from the Feynmann path integral which, in a sense, makes quantum states easier to visualize. Although we should admit that visualizing Feynmann path integral is still one level of complication above visualizing truly classical phenomena. 

The solution to the issue of nonlocality has the following ingredients. First of all, we made sure that the geometry is local, by replacing points with edges (Section \ref{Edges}). As it turns out, while the points have infinite number of neighbors, the edges have finite number of neighbors. That finite number, however, is controlled by a parameter $\Lambda$ that determines exactly what we mean by an edge-neighbor. Consequently, the action, partition function and probability density became $\Lambda$-dependent, which is why we wrote $S_{\Lambda}$, $Z_{\Lambda}$ and $\rho_{\Lambda}$. That $\Lambda$ is viewed as an integer-valued physical constant. Secondly, we made sure that the signals have finite lifetime by introducing the continuum measurement model (Sec \ref{EpsilonContinuum}) that lead to damping (Eq \ref{ContinuumAttenuation}). Consequently, whenever two edges happen to be remote from each other geometrically (thanks to $\Lambda$-based geometry just discussed) this, indeed, implies that their physical interaction is small. Thirdly, we introduced $\limsup$ and $\liminf$ technique to make sure that the theory remains mathematically well defined even if by some miracle the signals do not attenuate the way we expected (Sec \ref{StatesUnboundedMinimalist}) and, fourthly, we combined these three things to come up with a theory of quantum measurement on the unbounded causal set.

The main thing that makes the theory awkward is the use of $\limsup$ and $\liminf$. From Eq \ref{ContinuumAttenuation}, we have good physical reason to believe that the two coincide and, therefore can be replaced by simple $\lim$. However, we were unable to come up with a \emph{mathematical} proof of this \emph{physical} conjecture. So, for the purposes of this paper, we assumed the worst possible situation, where that conjecture isn't true, which is why we retained the use of $\limsup$ and $\liminf$. But, our goal for the future research, is to either prove or disprove the possibility of replacing $\limsup$ and $\liminf$ with a single $\lim$. 

For the purposes of making the theory as explicit as possible, this paper chose to use Spin-0 Lagrangian given in \cite{Delambertian1, Delambertian2}, with appropriate replacements of points with edges. However, the framework of this proposal is independent of the choice of Lagrangian that is being used. Therefore, it would work equally well if we replace the Lagrangians given in \cite{Delambertian1, Delambertian2} with, for example, Lagrangians given in \cite{TestFunctions}. By the same token, while this paper was focusing exclusively on scalar field, the framework can be adapted to include gauge fields. It would only require the replacement of scalar fields and Lagrangian used in this paper with the gauge fields and their Lagrangians that are given in, for example, in \cite{EMEdges} or \cite{TestFunctions}. While writing it down might take a few pages, conceptually it is a straightforward generalization of techniques presented in this paper. 

On a more mathematical front, some of the techniques of dealing with unbounded sets that are used in this paper might be similar to the ones used in nonstandard analysis. While there is a similarity in spirit, we did not pretend to develop full fledged formalism. It might be interesting, for the future research, to attempt to do so. In particular it might be interesting to see whether one can start from the concepts already developed in nonstandard analysis \cite{Nonstandard1, Nonstandard2} and manipulate them in such a way that would lead to the things we were using in this paper. We will leave this for future research.


\begin{thebibliography}{77} 

\bibitem{BoundaryTerms} Michel Buck, Fay Dowker, Ian Jubb, Sumati Surya ``Boundary Terms for Causal Sets" Class.Quant.Grav. 32 (2015) 20, 205004 DOI 10.1088/0264-9381/32/20/205004 and arXiv:1502.05388

\bibitem{BoundaryInfluence} William J. Cunningham ``Inference of Boundaries in Causal Sets" W.J. Cunningham, Inference of Boundaries in Causal Sets, Class. Quant. Grav. 35, 094002 (2018) DOI 10.1088/1361-6382/aaadc4 and arXiv:1710.09705

\bibitem{Mensky1} M.B. Mensky, “Quantum continuous measurements, dynamical role of information and restricted path integrals”, in Proceedings TH2002 (International Conference on Theoretical Physics) Supplement, Birkhauser 2003, and arXiv:quant-ph/0212112.

\bibitem{Mensky2} M.B. Mensky Quantum Measurement and Decoherence Kluwer Academic Publishers 2000

\bibitem{Kent}  A. Kent “Path integrals and reality” arXiv:1305.6565.

\bibitem{Epsilon} Roman Sverdlov, Luca Bombelli 1987 “Link between quantum measurement and the $i \epsilon$ term in the QFT propagator” Phys. Rev. D 90, 125020 (2014) and arXiv:1306.1948 

\bibitem{Review1} L Bombelli, J Lee, D Meyer and R Sorkin 1987 “Space-time as a causal set” Phys. Rev.
Lett. 59 521-524.

\bibitem{Review2} D.D. Reid; Introduction to causal sets: an alternate view of spacetime structure; Canadian Journal of Physics 79, 1-16 (2001); arXiv:gr-qc/9909075; (General)

\bibitem{Review3} F. Dowker, Introduction to causal sets and their phenomenology, Gen Relativ Gravit (2013) 45:1651–1667 doi:10.1007/s10714-013-1569-y

\bibitem{Review4} J. Henson, ``The causal set approach to quantum gravity", arXiv:gr-qc/0601121

\bibitem{Surya} S. Surya ``The causal set approach to quantum gravity", arXiv:1903.11544

\bibitem{Nowhere} Christian Wüthrich, Nick Huggett ``Out of Nowhere: Spacetime from causality: causal set theory" arXiv:2005.10873

\bibitem{Hawking} S W Hawking, A R King and P J McCarthy 1976 “A new topology for curved spacetime
which incorporates the causal, differential and conformal structures” J. Math. Phys. 17 174-181.

\bibitem{Malament} D Malament 1977 “The class of continuous timelike curves determines the topology of
spacetime” J. Math. Phys. 18 1399-1404.

\bibitem{CausalNonlocality1} Rafael Sorkin “Does locality fail at intermediate length-scales?” arXiv:grqc/0703099

\bibitem{CausalNonlocality2} ] L. Bombelli, J. Henson and R. Sorkin, Discreteness without Symmetry Breaking: a
Theorem, arXiv:gr-gc/060500v1

\bibitem{Delambertian1} R. Sorkin “Scalar Field Theory on a Causal Set in Histories Form” Journal of Physics:
Conference Series, Volume 306, Number 1

\bibitem{Delambertian2} F. Dawker, L. Glaser “Causal set d’Alembertians for various dimensions” J. Class.
Quant. Grav. (2013) IOP Publishing Ltd Classical and Quantum Gravity, Volume 30,
Number 19

\bibitem{CausalRecent} Fay Dowker, Nazireen Imambaccus, Amelia Owens, Rafael Sorkin and Stav Zalel ``A manifestly covariant framework for causal set dynamics" Classical and Quantum Gravity, Volume 37, Number 8

\bibitem{Percolation} F. Dowker, S. Surya; Observables in extended percolation models of causal set cosmology;Class. Quantum Grav. 23, 1381-1390 (2006); arXiv:gr-qc/0504069v1; (Cosmology, Dynamics, Observables)

\bibitem{Robb1}  A.A. Robb, \emph{A theory of time and space}, (Cambridge U. P., Cambridge, 1914)

\bibitem{Robb2} A.A. Robb, \emph{The absolute relations of time and space}, (Cambridge U. P. , Cambridge, 1921)

\bibitem{Robb3} A.A. Robb, \emph{Geometry of time and space}, (Cambridge U. P. , Cambridge, 1936). 

\bibitem{Nonstandard1} Alain Robert, \emph{Nonstandard Analysis} (Dover Publications, 2003)

\bibitem{Nonstandard2} Martin Vath, \emph{Nonstandard Analysis}  (Birkhauser; Text is Free of Markings edition, 2007)

\bibitem{GRW1} Ghirardi, G.C., Rimini, A., and Weber, T. (1986). "Unified dynamics for microscopic and macroscopic systems". Physical Review D. 34 (2): 470–491. Bibcode:1986PhRvD..34..470G. doi:10.1103/PhysRevD.34.470. PMID 9957165

\bibitem{GRW2} Ghirardi, Gian Carlo; Pearle, Philip; Rimini, Alberto (1990-07-01). "Markov processes in Hilbert space and continuous spontaneous localization of systems of identical particles". Physical Review A. 42 (1): 78–89. doi:10.1103/PhysRevA.42.78. PMID 9903779

\bibitem{RelativisticGRW} Roderich Tumulka ``A Relativistic Version of the Ghirardi-Rimini-Weber Model" J. Statist. Phys. 125 (2006) 821-840 DOI 10.1007/s10955-006-9227-3 and arXiv:quant-ph/0406094

\bibitem{GRWSverdlov} Roman Sverdlov ``Connection between GRW ``spontaneous collapse" and Mensky's ``restricted path integral" models" Foundations of Physics, 46(7), 825-835 2016 and arXiv:arXiv:1305.7516

\bibitem{QuantumLogic1} Rafael Sorkin ``Quantum Dynamics without the Wave Function" J.Phys.A40:3207-3222,2007 DOI 10.1088/1751-8113/40/12/S20 and arXiv:quant-ph/0610204

\bibitem{QuantumLogic2} H. Casini, ``The Quantum logic of causal sets" Class.Quant.Grav. 19 (2002) 6389-6404 DOI  10.1088/0264-9381/19/24/308 and arXiv:  gr-qc/0205013

\bibitem{EMEdges} R.Sverdlov ``Electromagnetic Lagrangian on a causal set that resides on edges rather than points" arXiv:1805.08064

\bibitem{TestFunctions} R.Sverdlov ``The use of test functions to help define quadratic Lagrangian on a causal set" arXiv:1807.07403 

\end{thebibliography}
\end{document}